\documentclass[letterpaper, 10 pt, conference]{ieeeconf}  

\IEEEoverridecommandlockouts                              

\usepackage[sorting=none, style=ieee]{biblatex}
\newtheorem{problem}{Problem}
\usepackage{hyperref}
\usepackage{graphics} 
\usepackage{epsfig} 
\usepackage{booktabs}
\usepackage{colortbl}
\usepackage{multirow}
\definecolor{mygray}{gray}{0.9}
\definecolor{mydarkgray}{gray}{0.85}
\definecolor{darkgray}{RGB}{169,169,169}
\definecolor{darkgray176}{RGB}{176,176,176}
\definecolor{gray}{RGB}{128,128,128}
\definecolor{green01270}{RGB}{0,127,0}

\usepackage{graphicx}
\usepackage{caption}
\usepackage{times} 
\usepackage{amsmath} 
\usepackage{amssymb}  
\usepackage[capitalise]{cleveref}
\usepackage{algpseudocode}
\usepackage{algorithm}

\def\IR{{\mathbb R}}

\title{\LARGE \bf
On the Design of Nonlinear MPC and LPVMPC for Obstacle Avoidance in Autonomous Driving*
}

\author{Maryam Nezami$^{1}$, Dimitrios S. Karachalios$^{1}$, Georg Schildbach$^{1}$ and Hossam S. Abbas$^{1}$ 
\thanks{*D. S. Karachalios is funded by the Deutsche Forschungsgemeinschaft (DFG, German Research Foundation) - 419290163.}
\thanks{$^{1}$Institute for Electrical Engineering in Medicine,
        University of Lübeck, Lübeck, Germany
        {\tt\small \{maryam.nezami, dimitrios.karachalios, georg.schildbach, h.abbas\}@uni-luebeck.de}}%
}

\begin{document}
\maketitle
\thispagestyle{empty}
\pagestyle{empty}
\begin{abstract}
In this study, we are concerned with autonomous driving missions when a static obstacle blocks a given reference trajectory. To provide a realistic control design, we employ a model predictive control (MPC) utilizing nonlinear state-space dynamic models of a car with linear tire forces, allowing for optimal path planning and tracking to overtake the obstacle.
We provide solutions with two different methodologies. Firstly, we solve a nonlinear MPC (NMPC) problem with a nonlinear optimization framework, capable of considering the nonlinear constraints. Secondly, by introducing scheduling signals, we embed the nonlinear dynamics in a linear parameter varying (LPV) representation with adaptive linear constraints for realizing the nonlinear constraints associated with the obstacle. Consequently, an LPVMPC optimization problem can be solved efficiently as a quadratic programming (QP) that constitutes the main novelty of this work. We test the two methods for a challenging obstacle avoidance task and provide qualitative comparisons. The LPVMPC shows a significant reduction in terms of the computational burden at the expense of a slight loss of performance. 
\end{abstract}

\section{INTRODUCTION}
In recent years, there has been a growing interest in developing autonomous driving vehicles. One of the key challenges in autonomous driving is navigating through complex environments and avoiding collisions with obstacles safely. Model predictive control (MPC) is a powerful control algorithm that has been widely used in the area of autonomous driving. MPC is particularly effective in controlling a vehicle because it can incorporate prior knowledge of the system dynamics, environmental information, as well as state and input constraints when computing a control input. Considering these factors, MPC can generate optimized control inputs that satisfy the constraints, resulting in high system performance and safety. 

MPC has been widely applied for obstacle avoidance in autonomous vehicles, see, e.g.,  \cite{nezami2021safe, nezami2022safe,brudigam2021stochastic}. It can be utilized to generate optimal trajectories that steer the vehicle away from obstacles in its path while respecting safety constraints. 
Given that vehicles are safety-critical systems, the use of nonlinear MPC (NMPC) is gaining popularity due to its ability to utilize high-fidelity nonlinear models of vehicle dynamics, thereby enabling more accurate and precise control actions.
The work presented in~\cite{allamaa2022real} has objectives similar to the current study, but it assumed constant longitudinal speed to solve the NMPC problem with sequential quadratic programs (SQP).
In~\cite{lee2022nonlinear}, an NMPC algorithm for path tracking has been proposed. This approach incorporates braking control before steering at high speeds. Investigating the effect of obstacle constraints on the algorithm's performance is interesting. This paper aims to keep the vehicle's operation stable while hard constraints are enforced in the optimization problem.
In~\cite{arrigoni2022mpc}, a method for generating safe and efficient driving trajectories for autonomous vehicles using NMPC has been introduced. The numerical solution of the NMPC was obtained using a genetic algorithm strategy, which does not offer a guarantee of convergence.   

The computational burden is a significant challenge for applying NMPC, particularly for systems with many states and constraints. Given the computational challenges, there has been increasing attention to linear parameter varying  (LPV) modeling methods to embed  nonlinear dynamics in a linear setting~\cite{de2022lpv, bujarbaruah2022robust}. Although the application of LPVMPC in autonomous driving has not yet received much attention in the literature, there are promising results reported in recent studies. In~\cite{nezami2022robust}, a control architecture for lane-keeping has been suggested where a tube-based LPVMPC  showed robust performance in lane-keeping. In~\cite{alcala2020lpv}, an online planning solution based on LPVMPC for autonomous racing has been proposed to improve the computational time while preserving the system's performance.  

\textit{Contributions}: 
This paper proposes a novel  LPV embedding for the nonlinear vehicle dynamics as a first step toward LPVMPC implementation. Such an LPV embedding  could be of interest  for convergence and feasibility analysis based on convex optimization tools. As a second step, at first, computation of the kinematic trajectories from a fixed map is presented. Then, we propose a linear formulation of the nonlinear constraints associated with the obstacle, which allows a tractable LPVMPC optimization problem using quadratic programming (QP). The proposed LPVMPC scheme integrates path planning and control into one optimization problem, deciding when to initiate the overtaking maneuver while ensuring the vehicle to be within the road boundaries. Finally, to verify the effectiveness of the proposed methods, simulation results are  compared to the full nonlinear implementation, and further discussions are given.

\textit{Contents}: \cref{sec:2} presents the nonlinear vehicle model and the nonlinear obstacle avoidance constraint, followed by the introduction of the linear representation of the obstacle constraint and the linear parameter varying modeling. In \cref{sec:3}, the models and constraints from \cref{sec:2} are used to set up the NMPC and the LPVMPC for obstacle avoidance. \cref{sec:4} shows the implementation  of the  methods and compares their performance in the proposed obstacle avoidance scenarios. Finally, a few concluding remarks are presented in \cref{sec:conc}.

\textit{Notations and definitions:} The notation $Q \succ 0$ represents the positive definiteness of a matrix $Q$. The weighted norm $\| x\|_Q$ is defined as $\| x\|_Q^2 = x^\top Q x $. The function $\text{diag}(\mathbf{x})$ constructs a diagonal matrix from a vector $\mathbf{x}$. A halfspace is defined as $\{ x \in \mathbb{R}^n | a^\top x \leq b\}$. The set of positive integers, including zero, is denoted by $\mathbb{Z}_{+} \cup \{ 0\}$.

\section{Vehicle Model and Constraints} \label{sec:2}
Consider the following discrete-time nonlinear system
\begin{equation}\label{eq:general_lin_model}
    z_{k+1} = f(z_k,u_k), \qquad \forall k \in \mathbb{Z}_{+} \cup \{ 0\},
\end{equation}
where $z_k \in \mathbb{R}^n$ and $u_k \in \mathbb{R}^m$ are the state and input vectors, respectively, at the instant $k$. The initial condition is $z_0$. The system is subject to the following state and input constraints:
\begin{equation}\label{eq:general_cons}
z_k \in \mathcal{Z}_k \quad \text{and} \quad u_k \in \mathcal{U} = \{ u_k \in \mathbb{R}^{m} | G^u u_k \leq h^u \}.
\end{equation}
Here, $\mathcal{Z}_k$ represents a time-varying set, and its formulation will be discussed in the subsequent section. Within the input constraint, we have $G^u \in \mathbb{R}^{q_u \times m}$ and $h^u \in \mathbb{R}^{q_u}$. The number of rows, $q_u$, depends on the number of inputs that have upper bounds, lower bounds, both upper and lower bounds, or no bounds at all.
In this section, the representation of the vehicle dynamics in the form \eqref{eq:general_lin_model}  using a dynamic bicycle model~\cite[p.~27]{rajamani2011vehicle} is given, as well as  the constraints formulation  to handle the obstacle avoidance.

\subsection{Dynamic Bicycle Model}\label{sec:bicycle_model}
\begin{figure}
    \centering
    \includegraphics[scale=0.6]{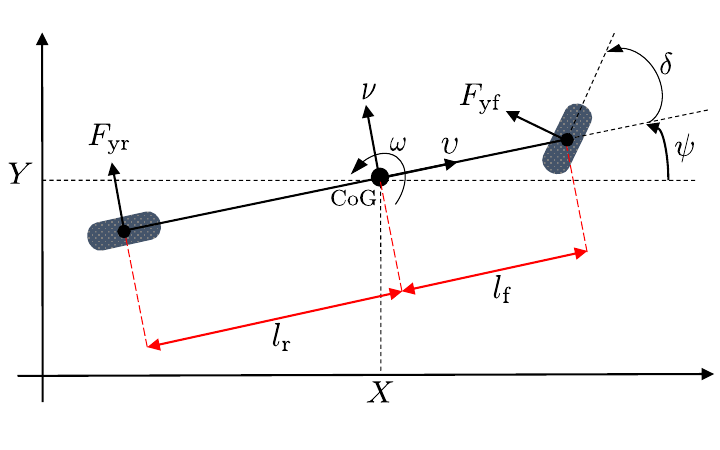}
    \caption{Vehicle dynamics representation}
    \label{fig:car_model}
\end{figure}

Based on~\cite[p.~27]{rajamani2011vehicle}, the differential equations describing the motion at time $t\geq 0$ of a vehicle are presented as follows
\begin{subequations}\label{eq:dynamic_model}
\begin{align}
    \dot{X}(t) &= \upsilon(t)\cos{\psi(t)}-\nu(t)\sin{\psi(t)},  \label{eq:Xdot}\\
    \dot{Y}(t) &= \upsilon(t)\sin{\psi(t)}+\nu(t)\cos{\psi(t)}, \label{eq:Ydot} \\
    \dot{\upsilon}(t) &= \omega(t) \nu(t) + a(t), \label{eq:xdotdot}\\
    \dot{\nu}(t) &= -\omega(t)\upsilon(t) + \frac{2}{m}( F_{\rm yf}(t) \cos{\delta(t)} + F_{\rm yr}(t)), \label{eq:ydotdot}\\
    \dot{\psi}(t) &= \omega, \label{eq:Psidot} \\
    \dot{\omega}(t) &= \frac{2}{I_{\rm z}} ( l_{\rm f} F_{\rm yf}(t) - l_{\rm r}  F_{\rm yr}(t)),\label{eq:psidotdot}
\end{align}
\end{subequations}
where $X$, $Y$, $\upsilon$, $\nu$, $\psi$ and $\omega$ denote the global $X$ axis coordinate of the center of gravity (GoG), the global $Y$ axis coordinate of the CoG, the longitudinal speed in body frame, the lateral speed in body frame, the vehicle yaw angle and the yaw angle rate, respectively. The control inputs of the system are the longitudinal acceleration $a$ and the steering angle $\delta$. The vehicle moment of inertia and mass are denoted by $I_{\rm z}$ and $m$, respectively. The lateral forces acting on the front and rear tires are denoted as $F_{\rm yf}$ and $F_{\rm yr}$, respectively, and  calculated as $F_{yf} = C_{\alpha \rm f}  \alpha_{\rm f}$, $F_{yr} = C_{\alpha \rm r}  \alpha_{\rm r}.$
The parameters $C_{\alpha \rm f}$ and $C_{\alpha \rm r}$ represent the cornering stiffness of the front and rear tire, respectively. The slip angle of the front tire is $\alpha_{\rm f}$ and is calculated as $\alpha_{\rm f} = \delta - (\nu + l_{\rm f} \omega)/\upsilon$. The rear tire slip angle is $\alpha_{\rm r}$ and is calculated as $\alpha_{\rm r} = (l_{\rm r} \omega -\nu )/\upsilon$. The parameters and variables are illustrated in~\cref{fig:car_model} and in~\cref{tab:VarPar}.

\begin{table}
    \centering
    \caption{Vehicle parameters}
    \label{tab:VarPar}
    \setlength{\tabcolsep}{2pt}
    \begin{tabular}{ccc} 
        \toprule
        \rowcolor{mydarkgray} \textbf{Symbol}   & \textbf{Variables} & \textbf{Unit} \\ 
                           $X$ & Global X-axis coordinates of the vehicle's CoG & m \\
        \rowcolor{mygray} $Y$ & Global Y-axis coordinates of the vehicle's CoG & m  \\
        $\upsilon$ & Longitudinal velocity of the vehicle & $\rm m/s$ \\
        \rowcolor{mygray} $\nu$ & Lateral velocity of the vehicle & $\rm m/s$ \\
         $\psi$ & Yaw angle of the vehicle & $\rm rad$  \\
       \rowcolor{mygray}  $\omega$  & Yaw rate of the vehicle & $\rm rad/s$  \\
         $\delta$ & Steering angle of the front tire & $\rm rad$ \\
       \rowcolor{mygray}  $a$ & Longitudinal acceleration of the vehicle & $\rm m/s^2$ \\
         $\alpha_{\rm f}$ & Front tire slip angle  & $\rm rad$ \\
       \rowcolor{mygray} $\alpha_{\rm r}$ & Rear tire slip angle  & $\rm rad$ \\ \toprule
	  \rowcolor{mydarkgray}    \textbf{Symbol}   & \textbf{Parameter} & \textbf{Value/Unit}  \\
         $C_{\alpha \rm f}$ & Cornering stiffness front tire & $156~\rm kN/rad$  \\
        \rowcolor{mygray}  $C_{\alpha \rm r}$ & Cornering stiffness rear tire & $193~\rm kN/rad$ \\
        $l_{\rm f}$ & Distance CoG to front axle & $1.04~\rm m$ \\
        \rowcolor{mygray} $l_{\rm r}$ & Distance CoG to rear axle & $1.4~\rm m$  \\
        $I_{\rm z}$ & Vehicle yaw inertia & $2937~\rm kgm^2$ \\
        \rowcolor{mygray} $m$ & Vehicle mass & $1919~\rm kg$  \\
    \end{tabular}
\end{table}

To utilize the model in~\cref{eq:dynamic_model} in an MPC framework, it is necessary to discretize the model. One of the commonly used methods for obtaining the corresponding  discrete-time system is the   forward Euler method\footnote{Forward Euler: $\dot{z}(t_k)\approx\frac{z(t_k+t_s)-z(t_k)}{t_s}$, for $t_k=t_sk,~k=0,1,\ldots$}. Therefore, 
the vehicle dynamics in~\cref{eq:dynamic_model} can be written as in~\cref{eq:general_lin_model},   
where $z_k = \begin{bmatrix} X_k & Y_k &\upsilon_k & \nu_k & \psi_k & \omega_k \end{bmatrix}^\top$, $u_k = \begin{bmatrix} \delta_k & a_k \end{bmatrix}^\top$ with the sampling time given in Table~\ref{tab:MPCPar}. 
\subsection{Constraints}\label{sec:bicycle_const}
To ensure the vehicle stays within the boundaries of the road, constraints are enforced on the $(X_k,Y_k)$ coordinates of the vehicle. One approach~\cite{tearle2021predictive}, involves computing the lateral error of the vehicle's center of gravity, $e_k^{\rm lat}$ as follows
\begin{equation}\label{eq:lateral_error}
    e_k^{\rm lat} = - \sin{(\psi_k^{\rm ref} )} ( X_k - X_k^{\rm ref}) + \cos{(\psi_k^{\rm ref})} (Y_k - Y_k^{\rm ref} ), 
\end{equation}
where $X_k^{\rm ref}$, $ Y_k^{\rm ref}$ and $\psi_k^{\rm ref}$ are the longitudinal position, the lateral position, and the orientation, respectively, on a point on a given reference trajectory at step $k$. Then, the following constraint ensures that the vehicle's CoG always stays within the boundaries of the road
\begin{equation}\label{eq:nonlinear:road_boudnry}
   - R_{1,k} \leq e_k^{\rm lat} \leq R_{2,k},
\end{equation}
where $R_{1,k}$ and $R_{2,k}$ are the road widths on the right and left sides of the reference trajectory at step $k$.  
The constraints associated with the road boundaries in \cref{eq:nonlinear:road_boudnry} can be computationally expensive due to their nonlinear nature. To address this challenge, an alternative linear constraint  is proposed as follows
\begin{equation}\label{eq:lin_road_boundry}
    \begin{bmatrix}
        a_{1,k} & b_{1,k} \\ 
        -a_{2,k} & -b_{2,k}
    \end{bmatrix} \begin{bmatrix}
        X_k \\ Y_k
    \end{bmatrix} \leq \begin{bmatrix}
        c_{1,k} \\ -c_{2,k}
    \end{bmatrix}, 
\end{equation}
where $-a_{1,k} X_k - b_{1,k} Y_k \geq -c_{1,k}$ and $ a_{2,k} X_k + b_{2,k} Y_k \geq c_{2,k}$ are  half-spaces defined by the tangent to the road boundary at step $k$, which ensure the $(X_k,Y_k)$ coordinates at step $k$  to remain within the two half-spaces. Imposing such linear constraints allows more efficient computations in the MPC optimization problem.

\begin{figure}
    \centering
    \includegraphics[scale= 0.3]{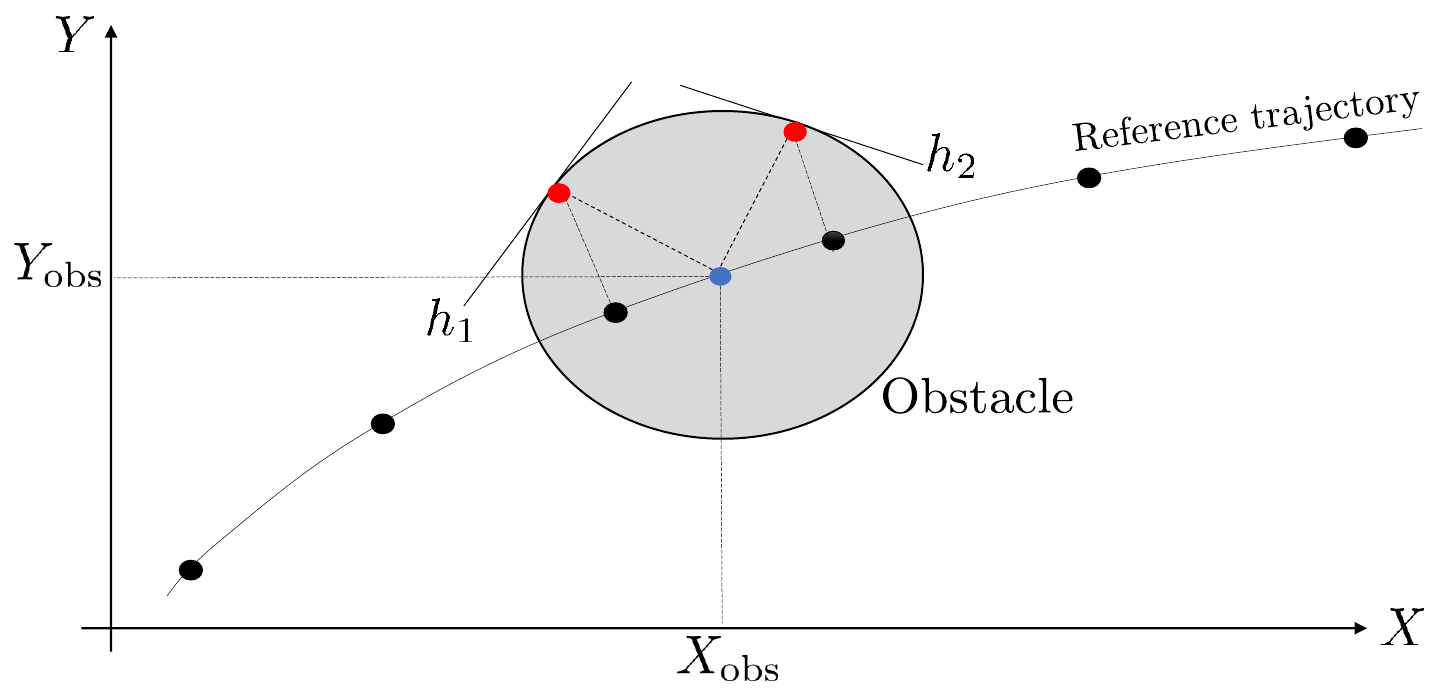}
    \caption{Linear obstacle constraint computation. The center of the obstacle $(X_{\text{obs}}, Y_{\text{obs}})$ is marked by the blue dot, and the reference trajectory is represented by the solid line with black dots.}
    \label{fig:lin_cons}
\end{figure}

An efficient representation of an  obstacle is to be formulated as an 
ellipse. For simplicity, we consider here circular obstacles.
To impose the obstacle constraints in the NMPC, one possible approach is to calculate the Euclidean distance between the $(X_k,Y_k)$ coordinates and the center of the obstacle and to ensure that the vehicle's $(X_k,Y_k)$ coordinates always remain outside the obstacle, as described below
\begin{equation}\label{eq:nonlinear_obs_cons}
    ( X_{\text{obs}} - X_k)^2 + (Y_{\text{obs}} - Y_k)^2 \geq r^2,
\end{equation}
where $(X_{\text{obs}}, Y_{\text{obs}})$ indicates the center of the obstacle and $r$ represents its radius. 
However, it is usually desired to formulate linear constraints for the obstacle in order to reduce the computational complexity.
For this purpose, we propose to replace the nonlinear constraint in~\cref{eq:nonlinear_obs_cons} with a linear inequality constraint, see~\cref{e:lin_cons}  below,
which varies over the MPC prediction horizon according to the tangent to the circular obstacle boundary. Every linear inequality constraint represents a  half-space, which defines a safe region to avoid collision with the  obstacle.
An illustration 
is depicted in Fig.~\ref{fig:lin_cons}. If a reference point falls inside the obstacle within the MPC horizon, a tangent defining a linear inequality constraint, such as $h_1$ or $h_2$, is calculated at the intersection point (the red dots in Fig.~\ref{fig:lin_cons}). The corresponding half-space includes the safe region to avoid the obstacle.
The  $(X_k,Y_k)$ coordinate of the vehicle should then be on that side, which can be defined by the linear inequality 
\begin{equation} \label{e:lin_cons}
    h_1 : a_{3,k} X_k + b_{3,k} Y_k \geq c_{3,k}, 
\end{equation}
where $a_{3,k}, b_{3,k}$ and $c_{3,k}$ are the parameters of the tangent half-space to the obstacle.  

\section{Controller Design}\label{sec:3}
\subsection{The Kinematic Trajectories From a Fixed Map}\label{sec:KinematicStates}
To consider a realistic setup for the problem, we assume that only the $(X_i^{\rm ref}, Y_i^{\rm ref})$ values of the reference trajectory are available. However, we should compute the corresponding reference values for the remaining four states, $\upsilon_k^{\rm ref}$, $\nu_k^{\rm ref}$, $\psi_k^{\rm ref}$ and $\omega_k^{\rm ref}$, to track the reference trajectory effectively. 
For the computation of $\psi_k^{\rm ref}$, the global $(X_i^{\rm ref}, Y_i^{\rm ref})$ can be directly used as follows
\begin{equation}\label{eq:psi_ref}
    \psi_k^{\rm ref} = \arctan\left(\frac{Y_{k-1}^{\rm ref} - Y_{k}^{\rm ref}}{X_{k-1}^{\rm ref} - X_{k}^{\rm ref}}\right). 
\end{equation}
Next, $\omega_k^{\rm ref}$ can be calculated as $\omega_k^{\rm ref} = (\psi_k^{\rm ref} - \psi_{k-1}^{\rm ref})/t_s$, where $\psi_{k-1}^{\rm ref}$ is the reference yaw angle which was computed in the previous step by using~\cref{eq:psi_ref}. To calculate $\upsilon_k^{\rm ref}$ and $\nu_k^{\rm ref}$, which represent the longitudinal and lateral speeds in the body frame, the reference points in the body frame are 
determined as follows 
\begin{equation}\label{eq:glob_to_loc}
    \begin{bmatrix}
        x_k^{\rm ref}  \\
        y_k^{\rm ref}
    \end{bmatrix} \!=\! \begin{bmatrix}
        \cos{(\psi_k^{\rm ref} )} & \sin{(\psi_k^{\rm ref}) } \\
        - \sin{(\psi_k^{\rm ref} )} & \cos{(\psi_k^{\rm ref} )}
    \end{bmatrix} \begin{bmatrix}
        X_{k}^{\rm ref} - X_{k-1}^{\rm ref}  \\ Y_{k}^{\rm ref} - Y_{k-1}^{\rm ref}
    \end{bmatrix}. 
\end{equation}
Then, 
the reference speeds
can be readily computed as $\upsilon_k^{\rm ref} = (x_k^{\rm ref} - x_{k-1}^{\rm ref})/t_s$ and $\nu_k^{\rm ref} = (y_k^{\rm ref}-y_{k-1}^{\rm ref})/t_s$, where $x_i^{\rm ref}$ and $y_i^{\rm ref}$ are computed in~\cref{eq:glob_to_loc}.

\subsection{Nonlinear MPC}
The constrained nonlinear optimal control for reference tracking w.r.t the decision variable 
$U = \{
    u_{0|k}, u_{1|k}, \ldots , u_{N-1|k} \}$
is formulated as follows.
 \begin{problem}[Nonlinear optimization problem]\label{Prob_NMPC}
 \vspace{4mm}
\begin{subequations}\label{eq:nonlinear_MPC}
	\begin{align} 
		\underset{U}{\text{min}} \
		& \| z_{N|k} \! -\! z^{\text{ref}}_{N|k} \|^2_P\! + \!\sum_{i=0}^{N-1} \| z_{i|k} - z^{\text{ref}}_{i|k} \|^2_Q +  \| u_{i|k}  \|^2_R   \\
		 \text{s.t.} \;\;
		& z_{i+1|k} \!= \!z_{i|k} \!+\! t_s f(z_{i|k},u_{i|k}),  \forall i \!= \!0,\ldots,N\!-\!1,  \label{eq:NMPC_model} \\
		& z_{0|k} = z_k, \label{eq:NMPC_initial_condition} \\
		&  z_{i|k} \in \mathcal{Z}_{i|k}, \quad \forall i = 0,1,\ldots,N,  \label{eq:NMPC_state_cons}\\
		&  u_{i|k} \in \mathcal{U}, \qquad \forall i = 0,1,\ldots,N-1, \label{eq:NMPC_input_cons}
	\end{align}
\end{subequations}
 \end{problem}
where $z^{\text{ref}}_{i|k} = \begin{bmatrix} X_{i|k}^{\rm ref} & Y_{i|k}^{\rm ref} & \upsilon_{i|k}^{\rm ref} & \nu_{i|k}^{\rm ref} & \psi_{i|k}^{\rm ref} & \omega_{i|k}^{\rm ref} \end{bmatrix}^\top$ is the reference value for the states at each step, which is computed by~\cref{eq:psi_ref,eq:glob_to_loc}. 
The tuning matrices are $Q \succeq 0 \in \mathbb{R}^{6 \times 6}$, $R \succ 0 \in \mathbb{R}^{2 \times 2}$ and $P \succeq 0 \in \mathbb{R}^{6 \times 6}$. The MPC prediction horizon is denoted with $N$. 
In the above optimization problem, $f$ is the nonlinear dynamic bicycle model in~\cref{eq:dynamic_model}, and $z_{0|k}$ is the system's initial condition. Here state constraint, $\mathcal{Z}_{i|k}$, includes the bounds on each state, the road boundary constraint~\eqref{eq:lin_road_boundry}, and the obstacle avoidance constraint~\eqref{eq:nonlinear_obs_cons}. The input constraint $\mathcal{U}$ is defined in~\cref{eq:general_cons}. 

\subsection{Linear Parameter Varying MPC}
By introducing the scheduling signals $\upsilon(t)$, $\nu(t)$, $\delta(t)$, and $\psi(t)$, that form the scheduling variable vector $p(t)=\left(\upsilon(t),~\nu(t),~\delta(t),~\psi(t)\right)$; the continuous-time nonlinear dynamics in~\cref{eq:dynamic_model} can be written equivalently
in the LPV representation as
\begin{equation}\label{eq:dis_model}
\left\{\begin{aligned}
    \dot{z}(t)&=A_c(p(t))z(t)+B_c(p(t))u(t),\\
    p(t)&=(\upsilon(t),\nu(t),\delta(t),\psi(t)),~t\geq 0.
    \end{aligned}\right.
\end{equation}
The state vector $z(t)$ of dimension $6$ can be defined as $z(t):=\left[\begin{array}{cccccc}
    X(t) & Y(t) & \upsilon(t) & \nu(t) & \psi(t) & \omega(t)\end{array}\right]^\top$ with initial conditions $z_0$ and the continuous-time system matrices $A_c\in\IR^{6\times 6},~B_c\in\IR^{2\times 2}$ as
\begin{equation*}
\small
\begin{aligned}
    A_c(p(t))&:=\left[\begin{array}{cccccc}
        0 & 0 & \cos(\psi(t)) & -\sin(\psi(t)) & 0 & 0 \\
        0 & 0 & \sin(\psi(t)) & +\cos(\psi(t)) & 0 & 0 \\
        0 & 0 & 0 & 0 & 0 & \nu(t) \\
        0 & 0 & 0 & a_{44}(t) & 0 & a_{46}(t) \\
        0 & 0 & 0 & 0 & 0 & 1\\
        0 & 0 & 0 & a_{64}(t) & 0 & a_{66}(t)\end{array}\right],\\
        \beta_f:=&\frac{2C{af}}{m},\beta_r:=\frac{2C{ar}}{m},~\gamma_f:=\frac{2\ell_fC_{af}}{I_z},\gamma_r:=\frac{2\ell_rC_{ar}}{I_z},\\
        a_{44}(t)&:=-\beta_f\cos(\delta(t))\frac{1}{\upsilon(t)}-\beta_r\frac{1}{\upsilon(t)},\\
        a_{46}(t)&:=-\upsilon(t)-\beta_f\cos(\delta(t))\frac{1}{\upsilon(t)}\ell_f+\beta_r\frac{1}{\upsilon(t)}\ell_r,\\
        a_{64}(t)&:=\frac{1}{\upsilon(t)}(\gamma_r-\gamma_f),~a_{66}(t):=-\frac{1}{\upsilon(t)}(\gamma_f\ell_f+\gamma_r\ell_r),
        \end{aligned}
\end{equation*}
and
\begin{equation*}
\small
    \begin{aligned}
       B_c(p(t)):= \begin{bmatrix}
           0 & 0 & 0 &  \beta_f\cos(\delta(t)) & 0 & \gamma_f \\
       0 & 0 &  1 &0 & 0 & 0 
       \end{bmatrix} ^\top.
    \end{aligned}
\end{equation*}
The discretization with the Euler method and a sampling time $t_s$ results in the discrete-time LPV representation of  \cref{eq:dis_model} as
\begin{equation}\label{eq:dis_LPV}
     \left\{\begin{aligned}      z_{k+1}&=A(p_k)z_k+B(p_k)u_k,\\
      p_k&=(\upsilon_k,\nu_k,\delta_k,\psi_k),~k\in\mathbb{Z}_+\cup \{ 0\}
      \end{aligned}\right.
\end{equation}
where $A(p_k)=I+t_s A_c(p_k),~B(p_k)=t_s B_c(p_k)$ are the discrete-time LPV system matrices, and $I\in\IR^{6\times 6}$ is the identity matrix.
\vspace{2mm}
\begin{problem}{QP optimization as $\texttt{QP}(p_{i|k},z_k,z_{i|k}^{\text{ref}})$}\label{Prob_LPVMPC}
\vspace{-2mm}
\begin{subequations}\label{eq:LPV_MPC}
	\begin{align} 
		\underset{U}{\text{min}} \
		& \| z_{N|k} \! -\! z^{\text{ref}}_{N|k} \|^2_P\! + \! \sum_{i=0}^{N-1}  \| z_{i|k} - z^{\text{ref}}_{i|k} \|^2_Q +  \| u_{i|k}  \|^2_R   \\
		 \text{s.t.} \;\;
		& z_{i+1|k}\!=\!\!A(p_{i|k})z_{i|k}\!\!+\!\!B(p_{i|k})u_{i|k},~i\!=\!0,\!\ldots\!,\!N\!\!-\!\!1 \label{eq:NMPC_model} \\
		& z_{0|k} = z_k, \label{eq:NMPC_initial_condition} \\
		& z_{i|k} \in \mathcal{\bar{Z}}_{i|k}, \quad \forall i = 0,1,\ldots,N,  \label{eq:NMPC_state_cons}\\
		& u_{i|k} \in \mathcal{U}, \qquad \forall i = 0,1,\ldots,N-1, \label{eq:LPVMPC_input_cons}
	\end{align}
\end{subequations}
\end{problem}
where, the reference trajectory $z^{\text{ref}}_{i|k}$, the tuning matrices $P$, $Q$ and $R$, as well as the decision variable $U$ are as defined for the Problem~\ref{Prob_NMPC}. The initial condition is~$z_{0|k}$, and $N$ denotes the MPC prediction horizon. The LPV model in~\cref{eq:NMPC_model} is defined in~\cref{eq:dis_LPV}. The input constraint $\mathcal{U}$ is given in~\cref{eq:general_cons}. The state constraint $\mathcal{\bar{Z}}_{i|k}$ in~\cref{eq:NMPC_state_cons} includes the bounds on the states, the road boundary constraint in~\cref{eq:lin_road_boundry} and the linear obstacle avoidance constraint in~\cref{e:lin_cons}. Therefore, the state constraint can be represented in the polytopic form $\mathcal{\bar{Z}}_k = \{ z_k \in \mathbb{R}^{6} | G^z_k z_k \leq h^z_k \}.$ The steps implementing the above LPVMPC are given in~\cref{alg:LPVMPC}. A similar algorithm has been proposed for the quasi LPV case in \cite{Cisneros_2016}.

\begin{algorithm}\label{algo:lpvmpc}
\caption{The QP-based LPVMPC algorithm}
\textbf{Input}: Initial conditions $z_k$, and the road reference $(X_k,Y_k),~k\in\mathbb{Z}_+$.\\
\textbf{Output}: The control input $u_k,~k=1,\ldots$, that drives the nonlinear system to the reference while avoiding obstacles.
\begin{algorithmic}[1]\label{alg:LPVMPC}
\State Initialize for $k=0$ the scheduling vector $\hat{p}_{i|0}$ as
$$\hat{p}_{i|0}:=\left(\upsilon_0,~\nu_0,~\delta_0,~\psi_0\right),~i=0,\ldots,N-1$$ 
\While{$k=0,1,\ldots$} 
\State Update the state $z_{i|k}^{\text{ref}}$ as explained in \cref{sec:KinematicStates}.
\State Solve the QP in Problem~\ref{Prob_LPVMPC}
\begin{equation*}
\begin{aligned}    
\left[z_{i+1|k},u_{i|k}\right]&\leftarrow\texttt{QP}(\hat{p}_{i|k},z_k,z_{i|k}^{\text{ref}}),~i=0,\ldots,N-1\\
\text{Update}~\hat{p}_{i|k}&:=\left(\hat{\upsilon}_{i|k},~\hat{\nu}_{i|k},~u_{i|k},~\hat{\psi}_{i|k}\right),~i=0,\ldots,N\\
\end{aligned}
\end{equation*}
  \State Apply $u_k=u_{0|k}$ to the system
  \State Measure $z_{k+1}$
  \State Update $\hat{p}_{i|k+1}=\hat{p}_{i+1|k},~i=0,\ldots,N-1$
  \State $k\leftarrow k+1$
  \EndWhile
\end{algorithmic}
\end{algorithm}

\section{Results and Discussions} \label{sec:4}
This section implements and compares the performance of the two MPCs, the NMPC and the LPVMPC. The simulations are performed on a Dell Latitude 5590 laptop with an Intel(R) Core(TM) i7-8650U CPU and 16 GB of RAM. The scenarios are implemented in Matlab~\cite{MATLAB:2019}, utilizing the YALMIP toolbox~\cite{Lofberg2004}, with an optimality tolerance of $10^{-4}$. To solve the nonlinear optimization problem, we employ the 
\textit{IPOPT} solver \cite{wachter2006implementation}. The Matlab \textit{quadprog} solver is used for solving the quadratic optimization problem. 

The simulation scenario is to  drive the vehicle in the middle on the right-hand side of a road to follow a reference trajectory using one of the proposed controllers in the previous sections. Then, an obstacle appears in the road, and the vehicle is controlled to overtake this obstacle safely and to return back to the reference trajectory in the middle on the right-hand side of the road. \cref{tab:MPCPar} presents the parameters utilized in the MPCs, along with the upper and lower limits of the states and input variables. 
\begin{table}[h]
    \centering
    \caption{MPC Parameters}
    \label{tab:MPCPar}
    \setlength{\tabcolsep}{2pt}
    \begin{tabular}{l l | l l} 
        \toprule
        \rowcolor{mydarkgray} \textbf{Parameter}   & \textbf{Value}  & \textbf{Parameter}    & \textbf{Value}\\ 
                       Lower bound on  $X_k$ &   -1 \text{m} & Upper bound on  $X_k$ &   150 \text{m}  \\
        \rowcolor{mygray} Lower bound on  $Y_k$ &   -1 \text{m} & Upper bound on  $Y_k$ &   120 \text{m} \\
        Lower bound on  $\upsilon_k$ &   1 $\frac{\rm m}{\rm s}$ & Upper bound on  $\upsilon_k$ &   100 $\frac{\rm m}{\rm s}$  \\
        \rowcolor{mygray} Upper bound on  $|\nu_k|$ &   10 $\frac{\rm m}{\rm s}$ & Upper bound on  $|\psi_k|$ &   $\pi/2 \text{ rad}$ \\
            Upper bound on  $|\omega_k|$ &   $\frac{\pi}{4t_s}\frac{\rm rad}{\rm s}$ & Sampling time $t_s$ & $0.05 \text{ s}$   \\
        \rowcolor{mygray}  Upper bound on  $| \delta_k |$ &   $\frac{34\pi}{180} \text{ rad} $ & Sampling frequency $f_s$ & $20 \text{ Hz}$\\
         Lower bound on  $a_k$ &   $-6$ $ \frac{ \rm m}{ \rm s^2}$ & Upper bound on  $a_k$ &   $2$ $ \frac{ \rm m}{ \rm s^2}$
    \end{tabular}
\end{table}

The reference trajectory is picked as a sine wave to mimic the road, and the $(X_i^{\rm ref}, Y_i^{\rm ref})$ on the reference trajectory are intentionally selected to be non-equidistant in space. As a result, the vehicle's speed shall be adjusted based on the distance between successive $(X_i^{\rm ref}, Y_i^{\rm ref})$ points. The initial condition of the vehicle is $z_0 = \begin{bmatrix} 0 & 0 & 10 & 0 & 0 & 0 \end{bmatrix}^\top$. Furthermore, the road on the left-hand side of the vehicle's reference trajectory is assumed to be always $4 \text{ m}$ wide, while on the right-hand side, it is always only $1 \text{ m}$ wide, i.e., $R_{1,k} = 1 \text{ m}$ and $R_{2,k} = 4 \text{ m}, \forall k = 0, 1, \dots$ in~\cref{eq:nonlinear:road_boudnry}. Also, for both MPCs, $N = 8$, $R = \text{diag}(\begin{bmatrix} 0.1 & 0.1\end{bmatrix})$, $Q = \text{diag}(\begin{bmatrix}10 & 10 & 1 & 1 & 1 & 1\end{bmatrix})$ and $P=Q$. To keep the vehicle movements smoother, we constrain  the rate of change of $\delta_k$ and $a_k$, i.e., $|\delta_{k} - \delta_{k-1} | \leq 40\pi / 180 \text{ rad}$, $| a_{k} - a_{k-1} | \leq 1.5$~$\rm m / \rm s^2$. 

To evaluate the effectiveness of our approach, we compare the performance of the NMPC and the LPVMPC in two problem setups: a reference tracking (RT) problem and an obstacle avoidance problem.

\subsection{Reference Tracking (RT)}
\begin{figure}
  \includegraphics[scale=0.9]{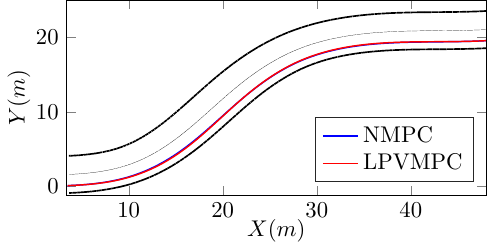}
  \caption{Reference tracking  by NMPC and LPVMPC }
    \label{fig:RT_states}
\end{figure}

\begin{figure}
\includegraphics[scale=0.9]{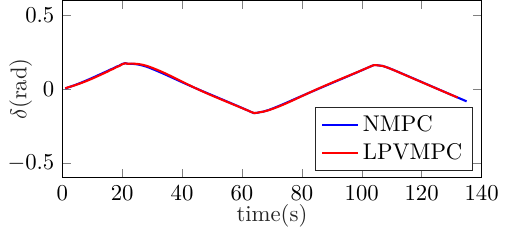}
  \caption{Reference tracking steering angles by NMPC and LPVMPC}
    \label{fig:RT_steering}
\end{figure}

\begin{figure}
  \includegraphics[scale=0.9]{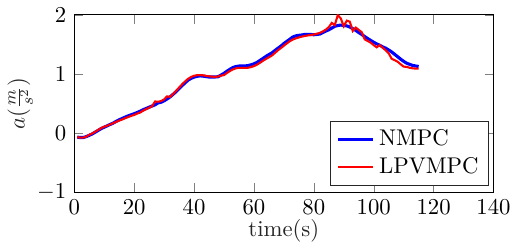}
  \caption{Reference tracking accelerations by NMPC and LPVMPC}
    \label{fig:RT_accel}
\end{figure}

In~\cref{fig:RT_states}, a comparison of the application of the NMPC and the LPVMPC for reference tracking is demonstrated. In this figure, the blue line represents the position of the vehicle when controlled by the NMPC, and the red line is the position of the vehicle controlled by the LPVMPC. 
As illustrated in~\cref{fig:RT_states}, the performance of the LPVMPC and the NMPC regarding tracking error is almost identical. By comparing the inputs generated by the controllers in~\cref{fig:RT_steering,fig:RT_accel}, it is clear that the steering angles produced by the NMPC and LPVMPC are nearly identical. Similarly, the accelerations are quite similar, although the NMPC acceleration appears to be smoother. \cref{tab:RT_Compuation_time} displays the computation time for solving an optimization problem to generate the inputs for the NMPC and the LPVMPC at each time instant $k$. The results confirm the reduction in the computation time   by using the LPVMPC. 
\begin{table}[h!]
    \centering
    \caption{Comparison of computation times using YALMIP between NMPC and LPVMPC for a reference tracking scenario}
    \label{tab:RT_Compuation_time}
    \setlength{\tabcolsep}{2pt}
    \begin{tabular}{ l | l | l} 
        \toprule
        \rowcolor{mydarkgray} \textbf{RT w/o obstacle}   & \textbf{NMPC}  & \textbf{LPVMPC}  \\ \hline
                              Average time &  0.2602 s &  0.0067 s \\
        \rowcolor{mygray}    Maximum time & 1.4996 s  & 0.0848 s \\
                             Minimum time  & 0.0739 s  & 0.0037 s   
    \end{tabular}
\end{table}

\subsection{Obstacle Avoidance}
In this subsection, the results of the comparison between the NMPC and the LPVMPC in an obstacle avoidance scenario are presented. The obstacle is assumed to be in a circular shape with a radius of $1 \text{ m}$ at $X_{\text{obs}} =  29.4819 \text{ m}$, $Y_{\text{obs}} = 17.4753 \text{ m}$. This means that the obstacle has blocked one side of the road in the studied scenario.

\begin{figure}
  \includegraphics[scale=0.9]{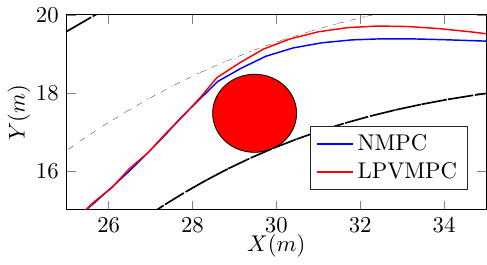}
  \caption{Obstacle avoidance scenarios by NMPC and LPVMPC. The car overtakes the obstacle with a longitudinal speed of around $50$ $\rm{km/h}$.}
  \vspace{-2mm}
    \label{fig:states}
\end{figure}

\begin{figure}
  \includegraphics[scale=0.9]{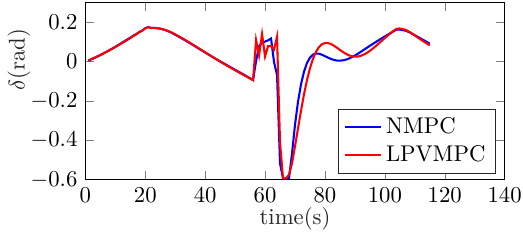}
  \caption{Obstacle avoidance steering angles by NMPC and LPVMPC}
    \label{fig:steering}
\end{figure}

\begin{figure}
  \includegraphics[scale=0.9]{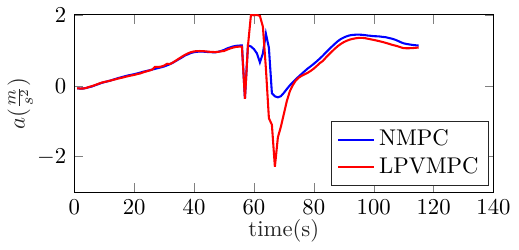}
  \caption{Obstacle avoidance accelerations by NMPC and LPVMPC}
    \label{fig:accel}
\end{figure}

The result of applying each of these MPCs to the nonlinear vehicle dynamics~\cref{eq:dynamic_model} is illustrated in~\cref{fig:states}. In this figure, the blue line represents the vehicle's trajectory when controlled by the NMPC, while the red line indicates its trajectory when controlled by the LPVMPC. As~\cref{fig:states} indicates, both the NMPC and LPVMPC are capable of controlling the vehicle to follow the desired reference trajectory and initiate the overtaking maneuver at an appropriate moment. However, the NMPC is performing a smoother maneuver. 
In~\cref{fig:steering,fig:accel}, the steering angles and the accelerations generated by each controller are presented. 
Based on the information in these figures, the NMPC can generate smoother control inputs. The difference in the inputs justifies the smoother movement of the vehicle in~\cref{fig:states}. 
The computation time of both the NMPC and the LPVMPC are presented in~\cref{tab:Compuation_time}. As the results confirm, using the LPVMPC can reduce the computation time significantly.
\begin{table}[h!]
    \centering
    \caption{Comparison of computation times using YALMIP between NMPC and LPVMPC for an obstacle avoidance scenario}
    \label{tab:Compuation_time}
    \setlength{\tabcolsep}{2pt}
    \begin{tabular}{ l | l | l} 
        \toprule
        \rowcolor{mydarkgray} \textbf{RT with obstacle}   & \textbf{NMPC}  & \textbf{LPVMPC}  \\ \hline
                              Average time &  0.7187 s &  0.0066 s \\
        \rowcolor{mygray}    Maximum time &  4.2031 s  & 0.0857 s \\
                             Minimum time  & 0.2559 s  & 0.0037 s   
    \end{tabular}
\end{table}

\section{CONCLUSIONS}\label{sec:conc}
This paper proposed a novel LPV embedding for modeling the nonlinear dynamics of a vehicle with linear tire forces. The proposed approach aims to simplify the implementation and offers a good alternative to NMPC, which is commonly considered. We introduced a linear formulation for obstacle avoidance constraints, enabling the proposed LPVMPC scheme to integrate both path planning and control into a single optimization problem. The LPVMPC is comparable to the NMPC in terms of performance with a more efficient computational burden. 
Finally, better tuning, model generalizations with the LPV embedding, dynamical and more challenging obstacles avoidance scenarios, 
as well as considering theoretical analysis such as  stability and recursive feasibility guarantees
are left for future research endeavors as the analysis with the LPV formulation can be carried out efficiently within the well-defined linear systems framework.


\printbibliography






\end{document}